\documentclass[twocolumn]{aastex61}

\usepackage{amsmath,graphics,graphicx}

\received{}
\accepted{}
\revised{}

\submitjournal{ApJL}

\shorttitle{ion velocity distributions at the base of the fast solar wind}
\shortauthors{Jeffrey et al.}

\begin{document}

	\title{Spectroscopic Measurements of the Ion Velocity Distribution at the Base of the Fast Solar Wind}
	\author{Natasha L. S. Jeffrey}
	\affil{School of Physics \& Astronomy, University of Glasgow, Glasgow, G12 8QQ, Scotland, UK}

	\author{Michael Hahn}
	\affil{Columbia Astrophysics Laboratory, Columbia University, MC 5247, 550 West 120th Street, New York, NY 10027, USA}

	\author{Daniel W. Savin}
	\affil{Columbia Astrophysics Laboratory, Columbia University, MC 5247, 550 West 120th Street, New York, NY 10027, USA}

	\author{Lyndsay Fletcher}
	\affil{School of Physics \& Astronomy, University of Glasgow, Glasgow, G12 8QQ, Scotland, UK}

	\correspondingauthor{Natasha L. S. Jeffrey}
	\email{natasha.jeffrey@glasgow.ac.uk}

\begin{abstract}
In situ measurements of the fast solar wind reveal non-thermal distributions of electrons, protons and, minor ions extending from $0.3$ AU to the heliopause. The physical mechanisms responsible for these non-thermal properties and the location where these properties originate remain open questions. Here we present spectroscopic evidence, from extreme ultraviolet spectroscopy, that the velocity distribution functions (VDFs) of minor ions are already non-Gaussian at the base of the fast solar wind in a coronal hole, at altitudes of $< 1.1 R_{\odot}$. Analysis of Fe, Si, and Mg spectral lines reveal a peaked line-shape core and broad wings that can be characteristed by a kappa VDF. A kappa distribution fit gives very small kappa indices off-limb of $\kappa\approx1.9-2.5$, indicating either (a) ion populations far from thermal equilibrium, (b) fluid motions such as non-Gaussian turbulent fluctuations or non-uniform wave motions, or (c) some combination of both. These observations provide important empirical constraints for the source region of the fast solar wind and for the theoretical models of the different acceleration, heating, and energy deposition processes therein. To the best of our knowledge, this is the first time that the ion VDF in the fast solar wind has been probed so close to its source region. The findings are also a timely precursor to the upcoming 2018 launch of the {\it Parker Solar Probe}, which will provide the closest in situ measurements of the solar wind at approximately $0.04$ AU ($8.5$ solar radii).
\end{abstract}


\section{Introduction}\label{intro}
Coronal holes, regions where the solar magnetic fields stretch far out into the heliosphere, are normally the source of the fast solar wind \citep[see][as a review]{2009LRSP....6....3C}. The fast solar wind displays significant non-equilibrium properties. In situ measurements at large heliospheric distances show ions flowing faster than electrons; and kappa distributions \citep[][]{1968ASSL...10..641O,1968JGR....73.2839V,2009JGRA..11411105L} of suprathermal particles are detected throughout the heliosphere \citep[cf.,][]{2006LRSP....3....1M}. For example, at $1$~AU the High Mass Resolution Spectrometer (MASS) on board {\em Wind} \citep{1995SSRv...71...79G} detected minor ions with high-energy tails in their speed distributions, which were well-described by a kappa function with $\kappa\approx2.5-4$ \citep{GRL:GRL9092}. Closer to the Sun, coronal hole spectroscopic observations have shown evidence for ion temperatures that are greater than the electron temperature \citep{Landi:ApJ:2009,2013ApJ...776...78H}, ion temperature anisotropies of $T_{\perp}/T_{\parallel}\approx10-100$  \citep{Kohl:ApJ:1998, 2009LRSP....6....3C, 2013ApJ...763..106H}, and non-Gaussian spectral line shapes at $\approx 2R_{\sun}$. These observations have been carried out using the Solar Ultraviolet Measurements of Emitted Radiation \citep[SUMER; ][]{1995SoPh..162..189W} and the Ultraviolet Coronagraph Spectrometer \citep[UVCS; ][]{1995SoPh..162..313K} on the {\em Solar and Heliospheric Observatory} \citep[{\em SOHO\,}; ][]{1995SSRv...72...81D}, and with the Extreme Ultraviolet (EUV) Imaging Spectrometer \citep[EIS; ][]{2007SoPh..243...19C} on board {\em Hinode} \citep{2007SoPh..243....3K}. 

In solar structures other than coronal holes, several spectroscopic studies have already inferred non-Gaussian spectral lines. \citet{2013arXiv1305.2939L} found evidence of non-Gaussian lines in solar active regions using EIS, while \citet{2017ApJ...842...19D} used the {\em Interface Region Imaging Spectrograph} \citep[{\em IRIS}; ][]{2014SoPh..289.2733D} to detect non-Gaussian lines in the transition region. \citet{2016A&A...590A..99J,2017ApJ...836...35J} found evidence of non-Gaussian spectral lines in different regions of two solar flares using EIS. \citet{2017ApJ...836...35J} showed that the physical part of the line could be well fitted with a kappa velocity distribution function (VDF) giving $\kappa < 10$, suggestive of non-Maxwellian ion VDFs. By fitting a convolution of different instrumental and physical profiles, they found that the large range of $\kappa$ produced for a given line width made a physical rather than instrumental origin for these features more likely.

Non-thermal processes are expected to be more evident in coronal holes where the density is lower and collisions that thermalize the plasma, are rarer. Indeed, many non-thermal properties of the solar wind likely originate in the corona rather than developing locally. Analyses of in situ data have shown that some non-thermal properties in the solar wind are reduced in parcels of wind that have undergone greater collisional relaxation between the Sun and the observation point \citep{2002ESASP.508..361C, Kasper:PRL:2008}. This suggests that the underlying processes producing non-thermal properties take place near the Sun. Furthermore, non-Gaussian spectral line shapes could be produced directly by the fluid motions \citep[waves or turbulence, e.g.,][]{2004physics..12091M,2006CoPP...46..672M} that pervade the solar wind \citep[cf., ][]{1995SSRv...73....1T,2005LRSP....2....4B,2013SSRv..178..101A} and affect its generation and evolution.

Here using spectral lines observed with EIS, we provide evidence that non-Gaussian VDFs exist at low altitudes ($<1.1R_{\sun}$) in a polar coronal hole at the base of the fast solar wind. These data provide important observational constraints for models of solar wind heating, acceleration, and energy deposition. For this study, we analyzed the line shapes of five different ions as we describe below.

\section{Data Analysis}\label{data_analysis}

\begin{figure*}[t]
\centering
\includegraphics[width=0.49\linewidth]{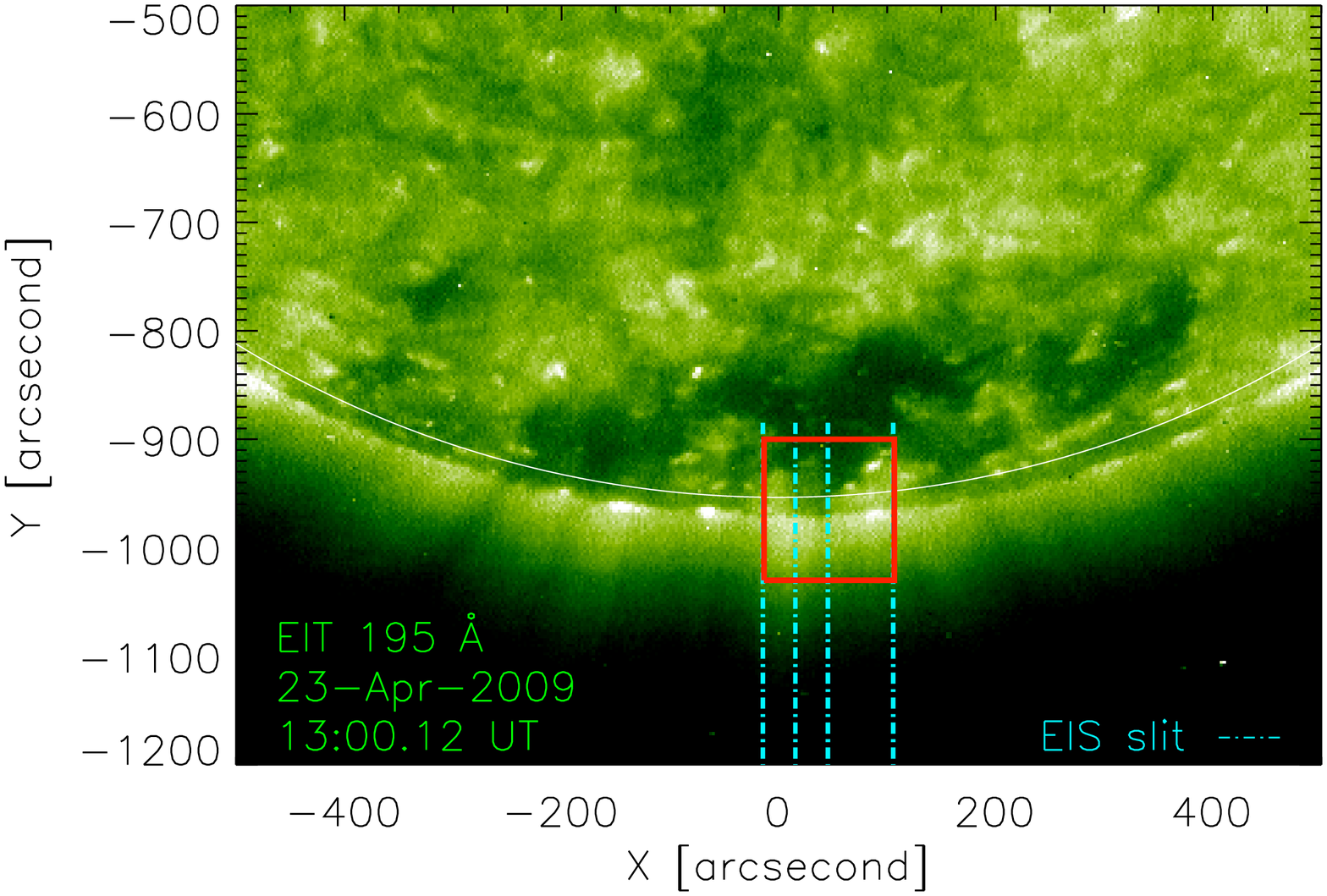}
\includegraphics[width=0.4885\linewidth]{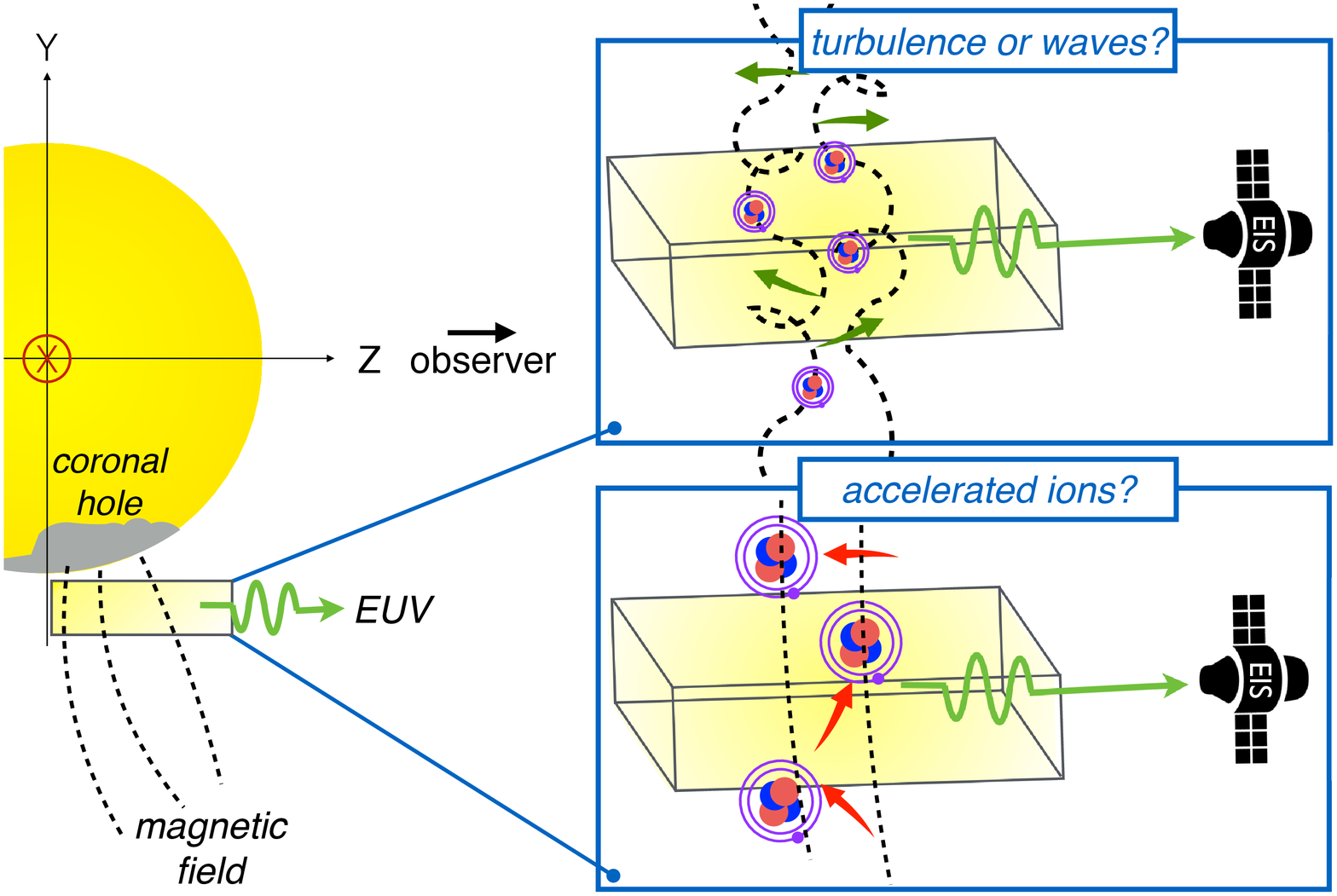}
\caption{{\it Left:} An EIT 195 \AA~image of the coronal hole. The four EIS slit locations used are shown. The red rectangle denotes the range of $Y$ values found to be suitable for the analysis ($\approx -900\arcsec>Y>-1030 \arcsec$). The solar limb at $R_{\odot}\approx954 \arcsec$ is denoted by the white curve. {\it Right: } A cartoon of the observation and the possible processes producing non-Gaussian line shapes. $Z$ denotes the line-of-sight direction.}
\label{fig1}
\end{figure*}

We analyzed EUV spectral data from the southern polar coronal hole observed by EIS on 23 April 2009. A {\em SOHO} Extreme ultraviolet Imaging Telescope \citep[EIT; ][]{1995SoPh..162..291D} 195 \AA~image of the coronal hole is shown in Figure \ref{fig1}. The EIS 2$^{\prime\prime}$ slit locations for four separate thirty-minute observations are shown. The data from each were combined for this analysis. The data preparation and averaging have been previously described in \citet{2012ApJ...753...36H} and \citet{2013ApJ...776...78H}. We also binned the data along the slit in order to increase the signal-to-noise ratio. We used a $15 \arcsec$ binning for \ion{Si}{7} and \ion{Fe}{8}, $20 \arcsec$ for \ion{Fe}{9}, and 25$\arcsec$ for \ion{Mg}{7} and \ion{Si}{10}. Additionally, we accounted for the EIS spectral pixel size of $\Delta \lambda=0.022$~\AA~using the code icsf.pro \citep{2016SoPh..291...55K}, which accounts for the finite $\lambda$ binning.

Here, we apply the EIS spectral line shape analysis of \citet{2016A&A...590A..99J,2017ApJ...836...35J} to these coronal hole data. We focus on unblended spectral lines and fit the lines with three different functions: (1) a single Gaussian (SG), (2) a convolved kappa-Gaussian (KG) and (3) a double Gaussian (DG). The KG function $\mathcal{W}(\lambda)$ is a convolution of a Gaussian $\mathcal{G}$ and a kappa function $\mathcal{K}$, where $\lambda$ is the wavelength. It has five free fit parameters labeled $A[j]$: 
\begin{equation}
\begin{split}
&\textstyle \mathcal{W}(\lambda) = \mathcal{G}(\lambda)*\mathcal{K}(\lambda) = A[0]+A[1]\times\\ 
&\textstyle \sum_{\lambda^{'}}\Bigg\{\exp{\left(-\frac{(\lambda^{'}-A[2])^{2}}{2\sigma^{2}(Y)}\right)\Bigg\} \Bigg \{\left(1+\frac{(\lambda-\lambda^{'}-A[2])^{2}}{2A[3]^{2}A[4]}\right)^{-A[4]+1}} \Bigg \}.
\end{split}
\label{fit_kG}	
\end{equation}
The Gaussian function in the first pair of curly brackets accounts for the EIS instrumental broadening, which is characterized by a standard deviation $\sigma$. For a given EIS slit width, $\sigma$ is a function of $Y$ along the EIS charge-coupled device (CCD). The second pair of curly brackets contains the kappa function, which parameterizes any non-Gaussian properties of the line shape. Lines with $\kappa \gtrsim 20$ are indistinguishable from Gaussian and lines with $\kappa < 10$ show clearly enhanced wings. The kappa line profile in Equation (\ref{fit_kG}) is derived from a three-dimensional kappa distribution of the first kind integrated over velocities perpendicular to the line-of-sight \citep{2017ApJ...836...35J}, and the line-of-sight VDF is $\propto I(\lambda)\frac{d\lambda}{dv_{\parallel}}=I(\lambda)\frac{\lambda_{0}}{c}$, where $I$ is the line intensity, $v_{\parallel}$ the line-of-sight velocity, $\lambda_{0}$ the rest wavelength and $c$ the speed of light.

Our use of the kappa line profile is primarily a convenience for detecting departures from Gaussian and does not necessarily mean that the underlying ion or plasma VDF is precisely a kappa distribution. The double Gaussian (DG) function checks whether a two-component fit is sufficient to describe line profiles that cannot be adequately fitted using a single Gaussian. A two-component line profile could indicate that a component of the plasma has a bulk flow velocity or it might indicate the presence of two distinct structures with different characteristic profiles along the line-of-sight \citep[e.g.,][]{Chae:ApJ:1998}.

\begin{table*}[t]
\begin{center}
\caption{The properties and fitting results of the lines studied. $T$ indicates the formation temperature taken from the CHIANTI line list \citep{1997A&AS..125..149D,2013ApJ...763...86L}, $Q/M$ is the charge-to-mass ratio, and $\kappa$ the kappa index. The table lists the ions in the order of their formation temperature, weakly suggesting that $\kappa$ increases with temperature.}
\begin{tabular}{c c c c c c c c}
\hline\hline
Ion & Wavelength (\AA) & $\log T$ & Q/M & No. of altitudes & Full $\kappa$ range & Off-limb $\kappa$ range & Off-limb $\left<\kappa\right>$\\ \hline
\ion{Fe}{8} & 186.599 &5.7 & 0.14 & 7 &$1.6-2.3$ &$1.9-2.3$& 2.1\\ \hline
\ion{Si}{7} & 275.361 &5.8 & 0.25 &  8 &$1.8-2.4$ &$2.1-2.4$&2.3\\ \hline
\ion{Mg}{7} & 276.154 &5.8 & 0.29 & 1 & 2.1& 2.1 &2.2\\ \hline
\ion{Fe}{9} & 197.862 &6.0 & 0.16 & 6 & $2.0-2.3$ & $2.2-2.4$ &2.3\\ \hline
\ion{Si}{10} & 258.374 &6.2 & 0.36 & 1 & 2.6 &2.6 & 2.6\\ \hline
\end{tabular}
\label{tb1}
\end{center}
\end{table*}
\begin{figure*}[t]
\centering

\includegraphics[width=0.37\linewidth]{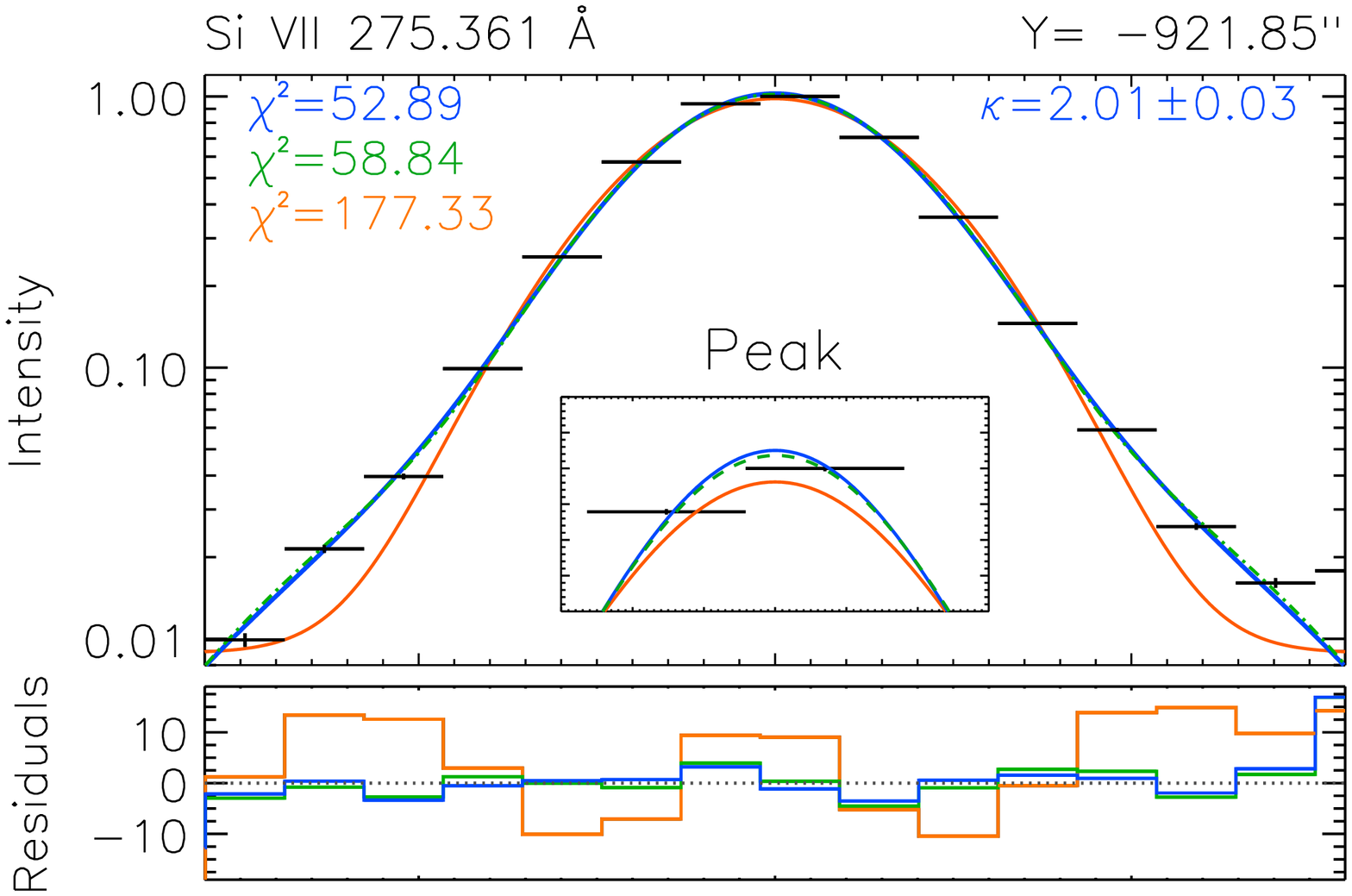}
\hspace{-39pt}
\includegraphics[width=0.37\linewidth]{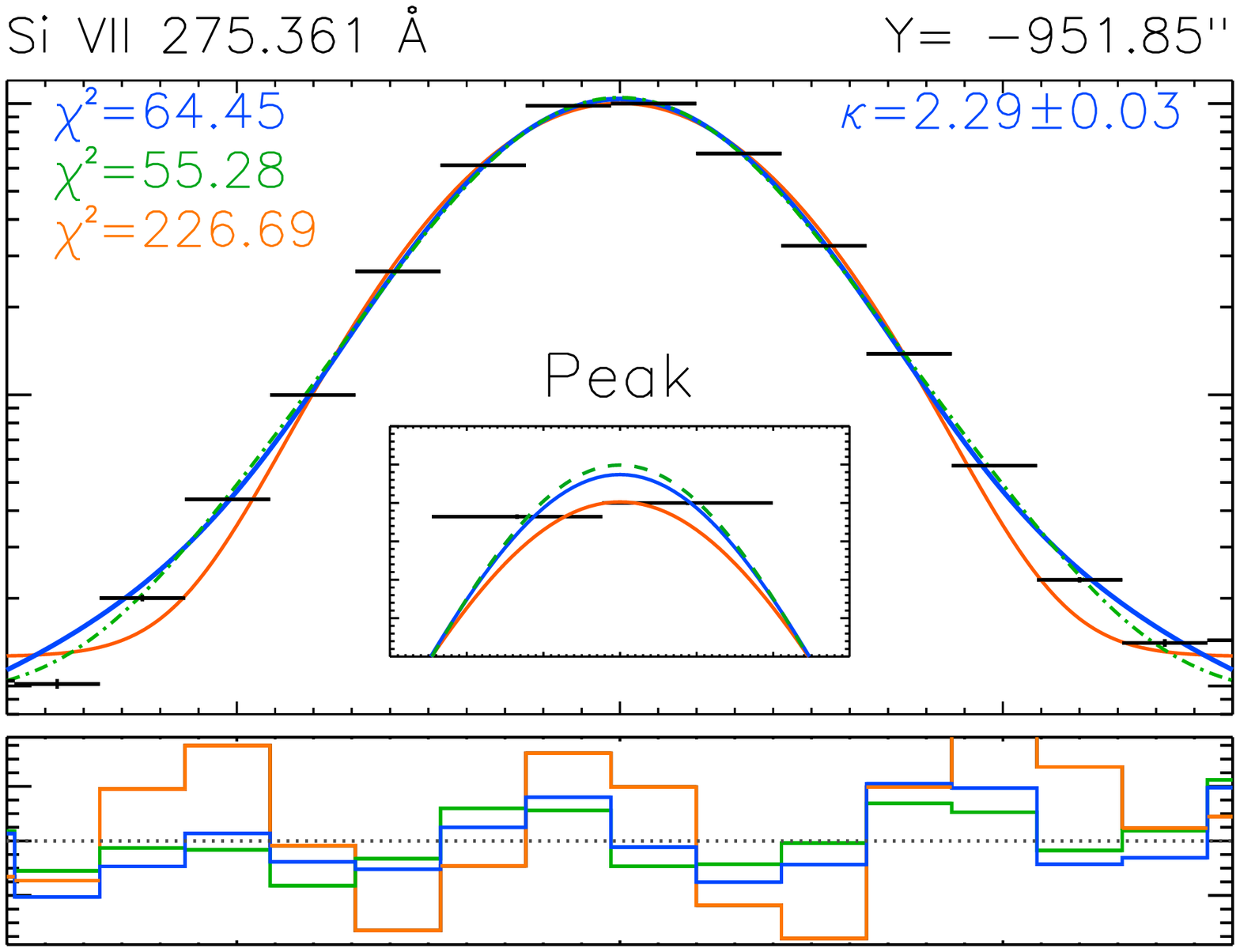}
\hspace{-39pt}
\includegraphics[width=0.37\linewidth]{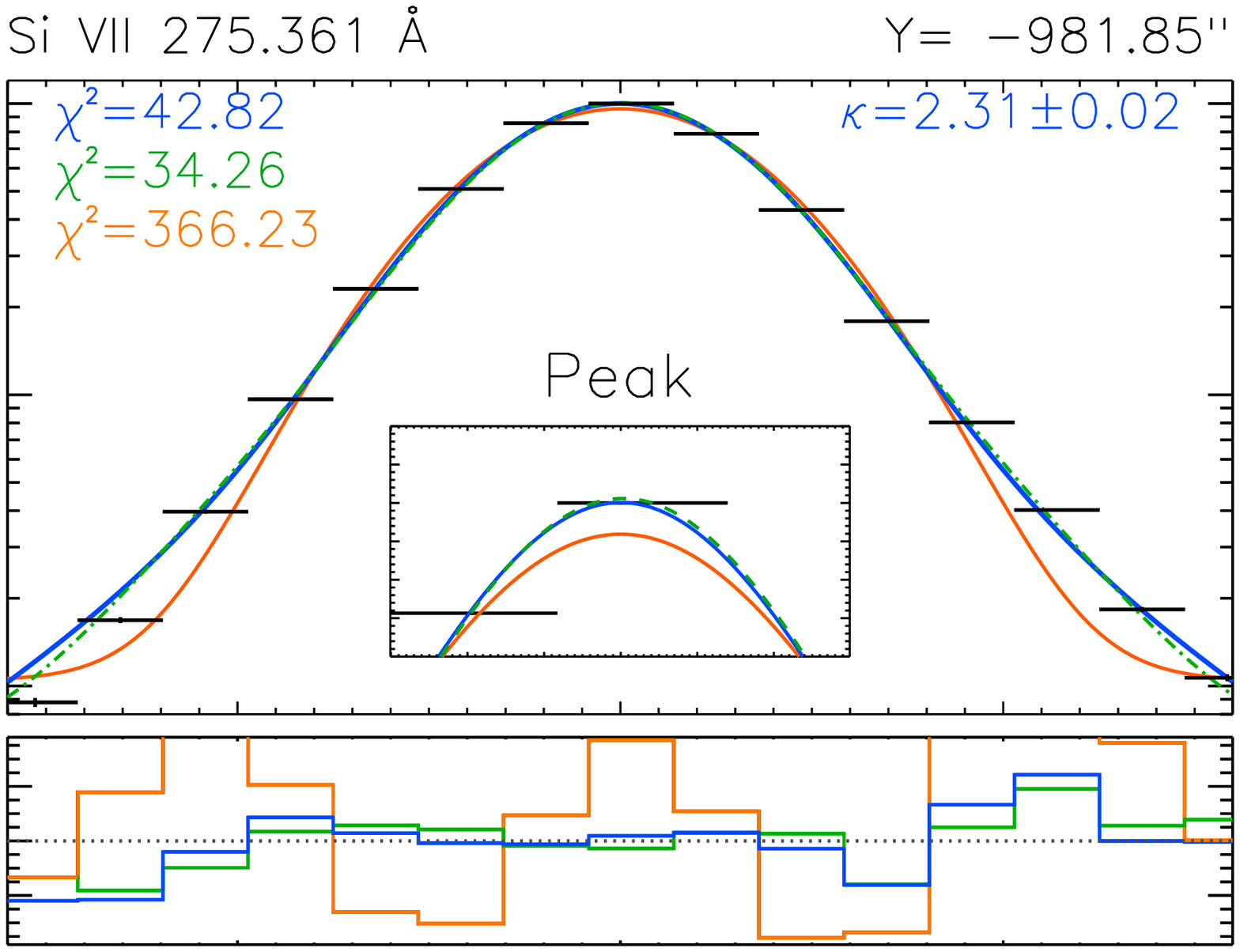}

\vspace{-15pt}

\includegraphics[width=0.37\linewidth]{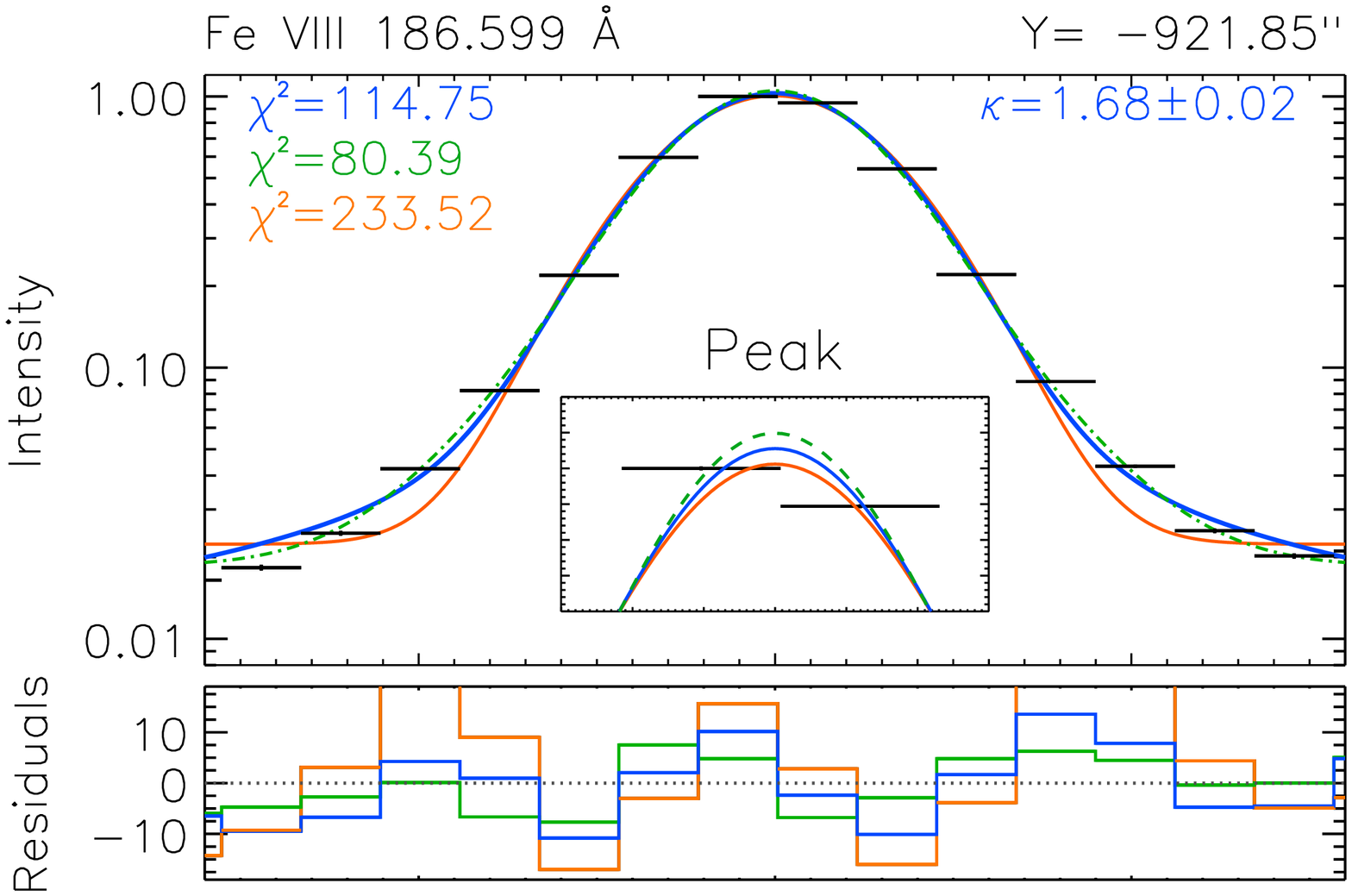}
\hspace{-39pt}
\includegraphics[width=0.37\linewidth]{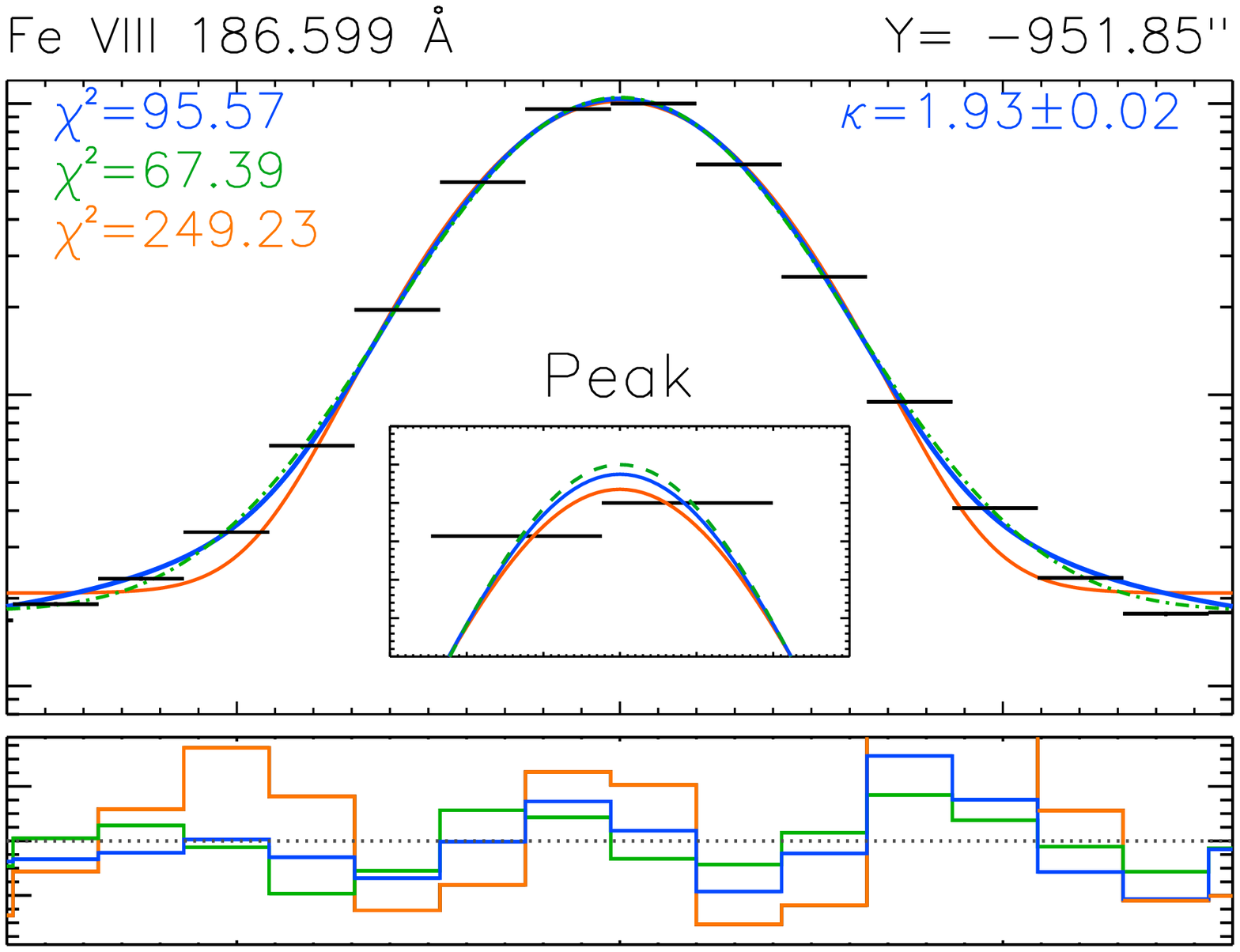}
\hspace{-39pt}
\includegraphics[width=0.37\linewidth]{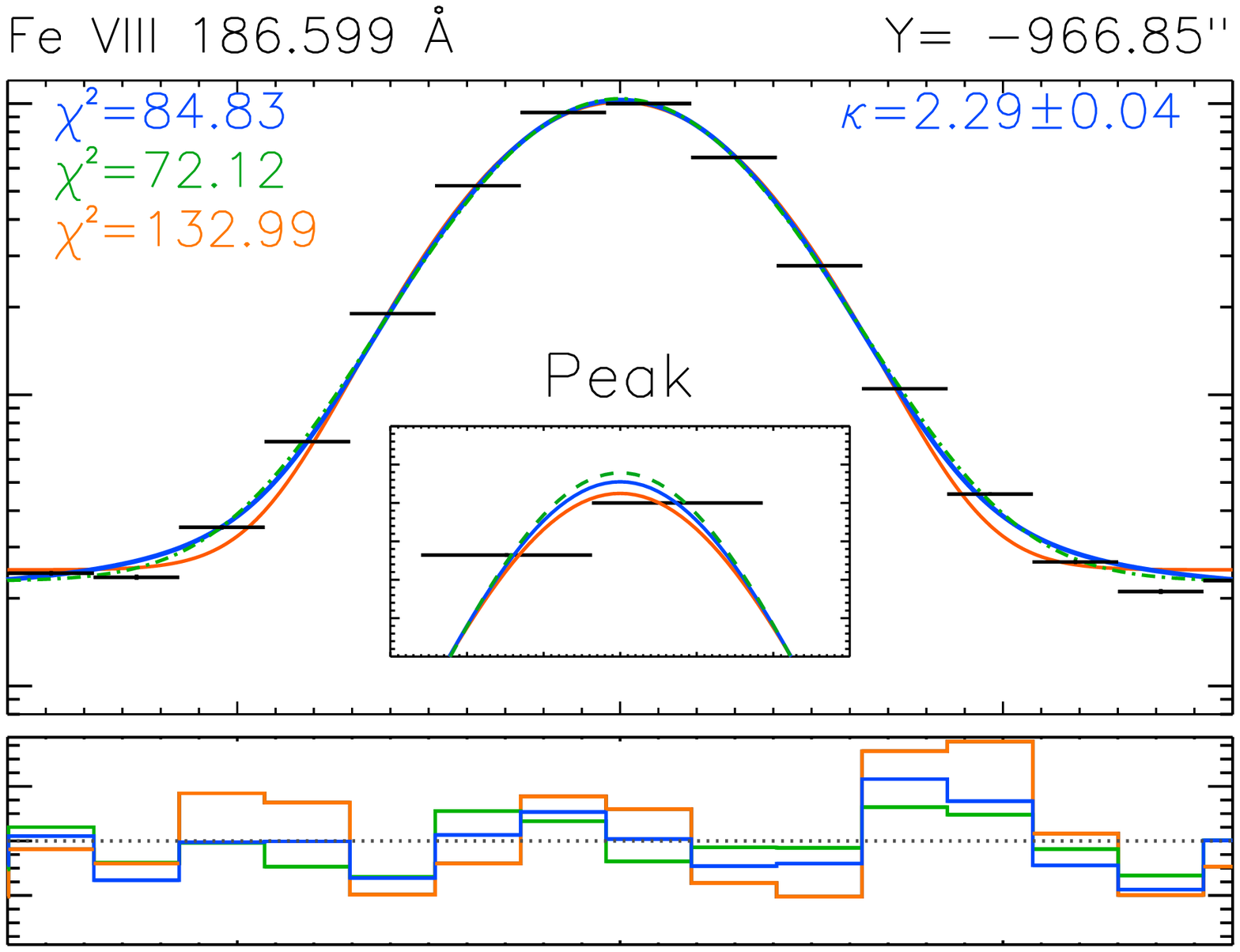}

\vspace{-15pt}

\includegraphics[width=0.37\linewidth]{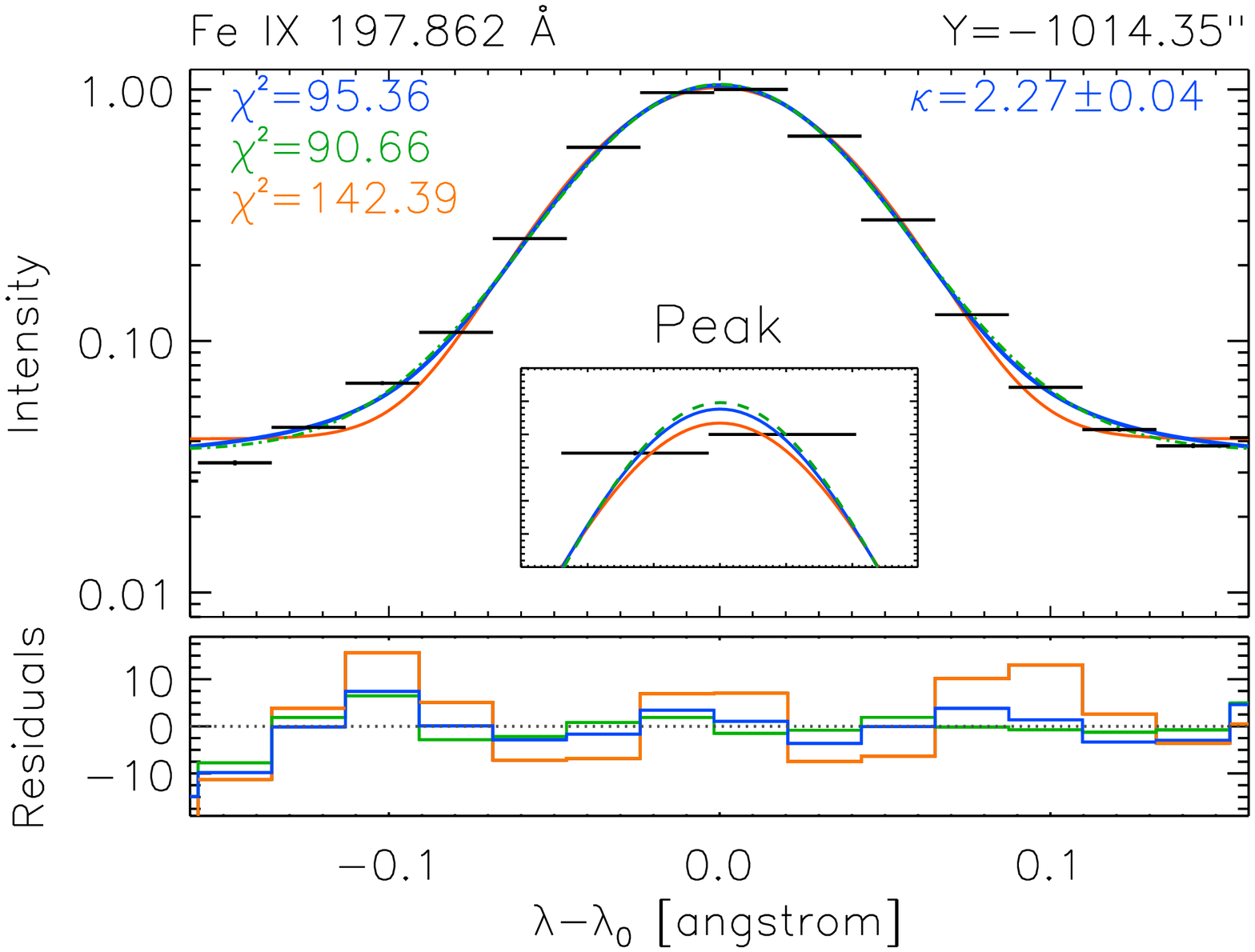}
\hspace{-39pt}
\includegraphics[width=0.37\linewidth]{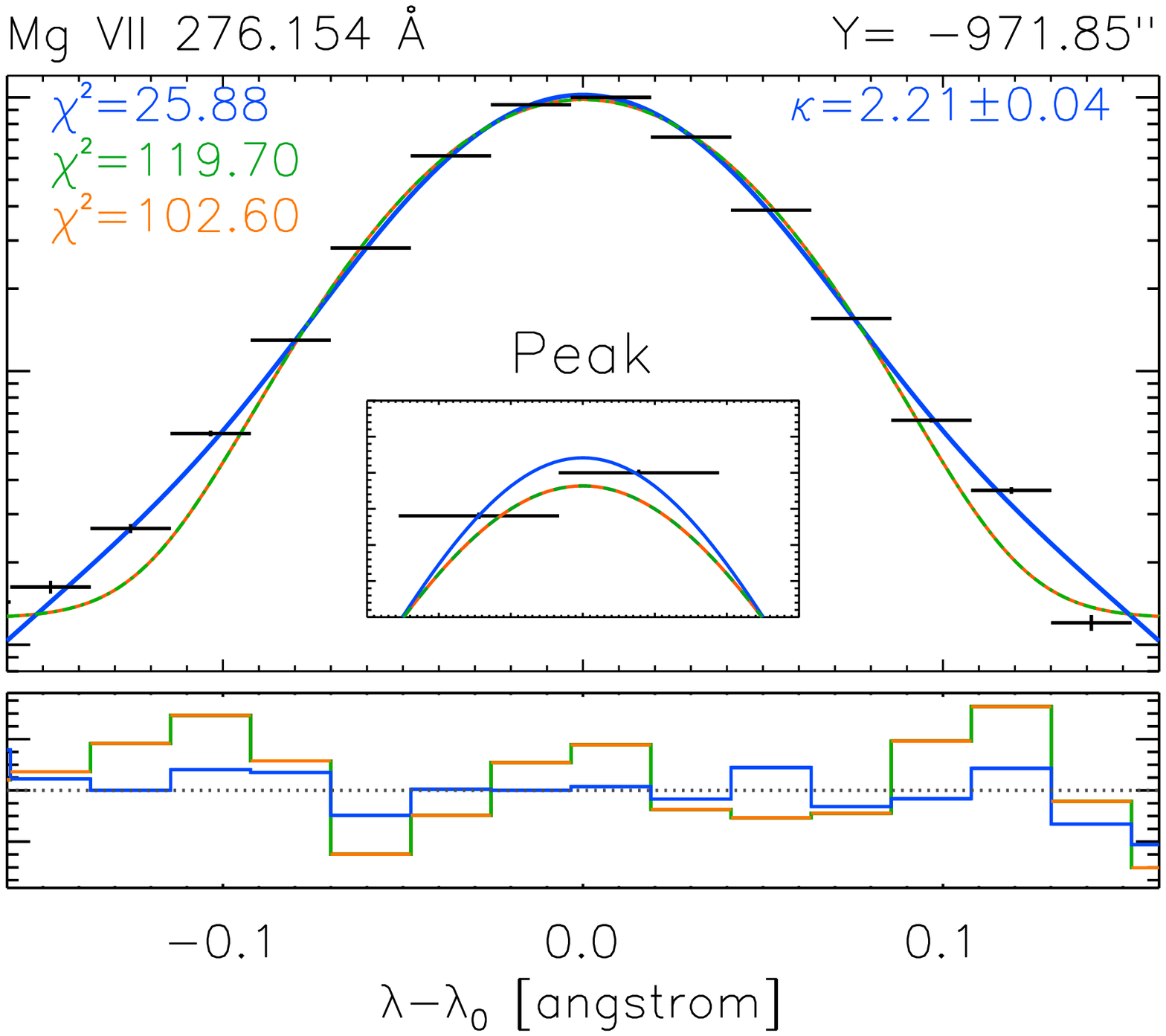}
\hspace{-39pt}
\includegraphics[width=0.37\linewidth]{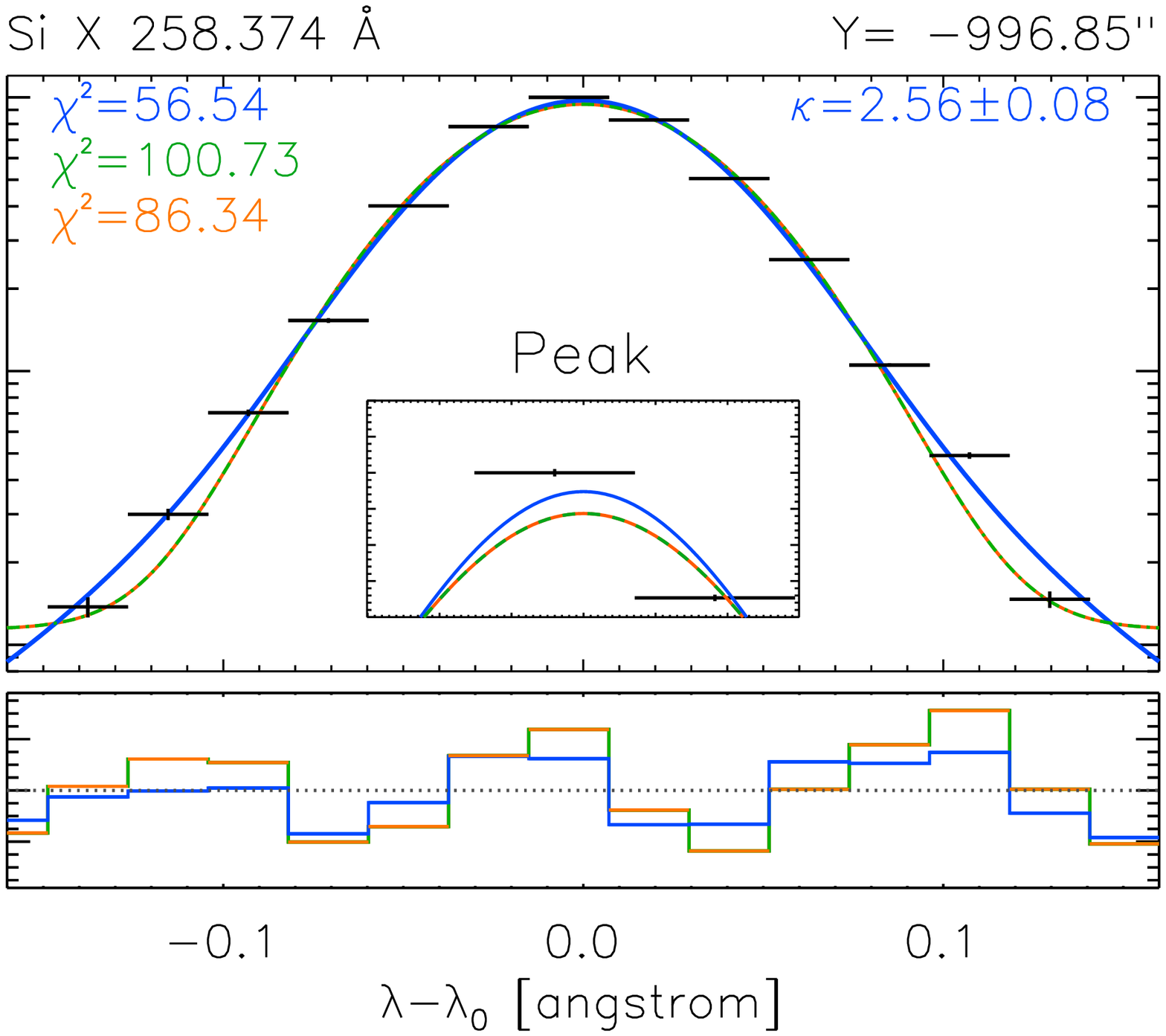}

\caption{Line fitting results for five lines using the spatial binning described in Section \ref{data_analysis}. {\it Top row -} \ion{Si}{7} at three altitudes, {\it middle row -} \ion{Fe}{8} at three altitudes, and {\it bottom row -} one example of \ion{Fe}{9} {\it (left)}, \ion{Mg}{7} {\it (middle)}, and \ion{Si}{10} {\it(right)}. Each panel shows SG (orange), DG (green), and KG (blue) fits, as well as the $\chi_{r}^{2}$ values and the $\kappa$ indices determined from the KG fits. The horizontal bars represent the EIS spectral pixel size of $\Delta \lambda = 0.022$~\AA.}
\label{fig2}
\end{figure*}

\begin{figure*}[t]
\centering
\includegraphics[width=0.497\linewidth]{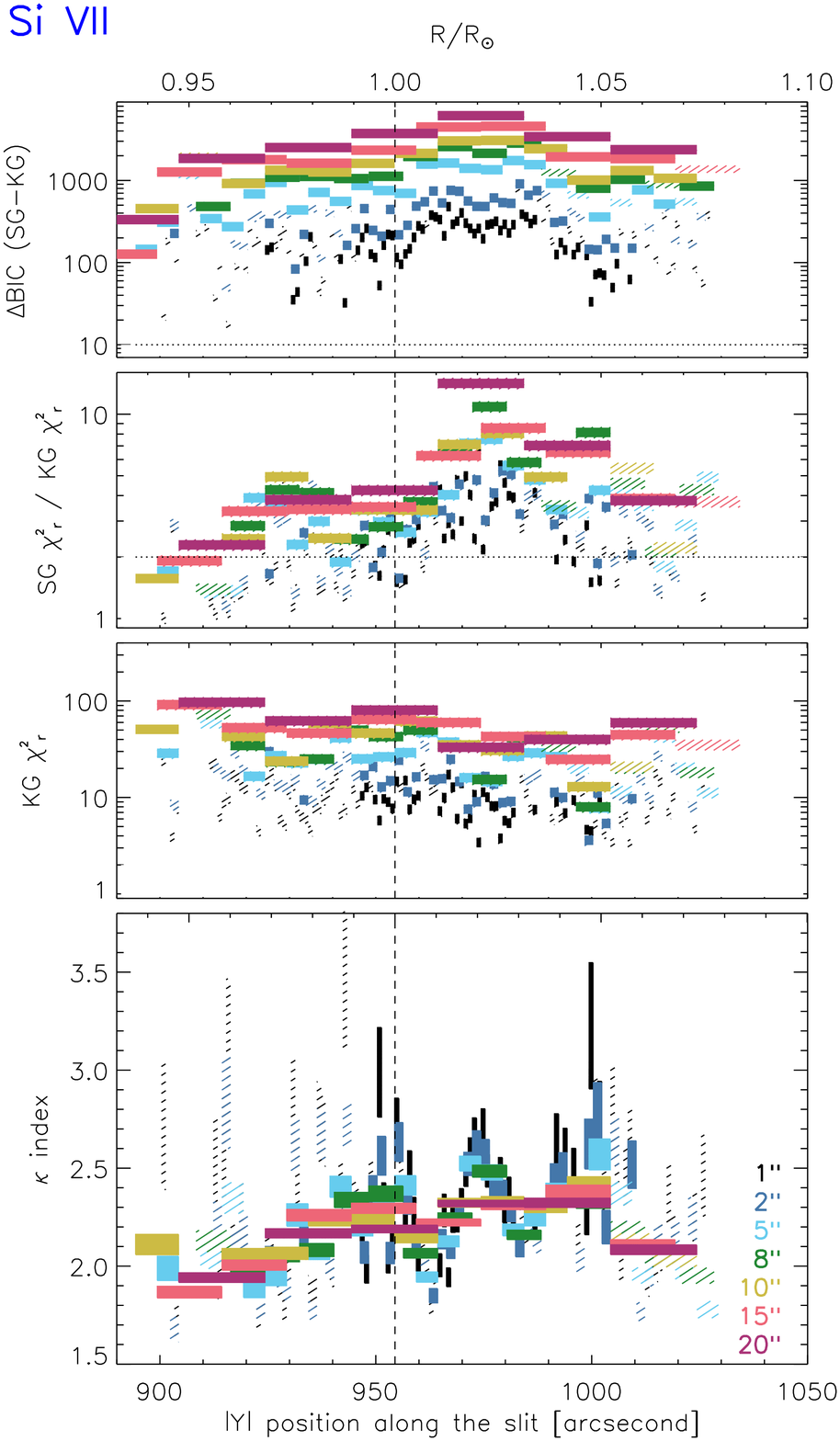}
\includegraphics[width=0.497\linewidth]{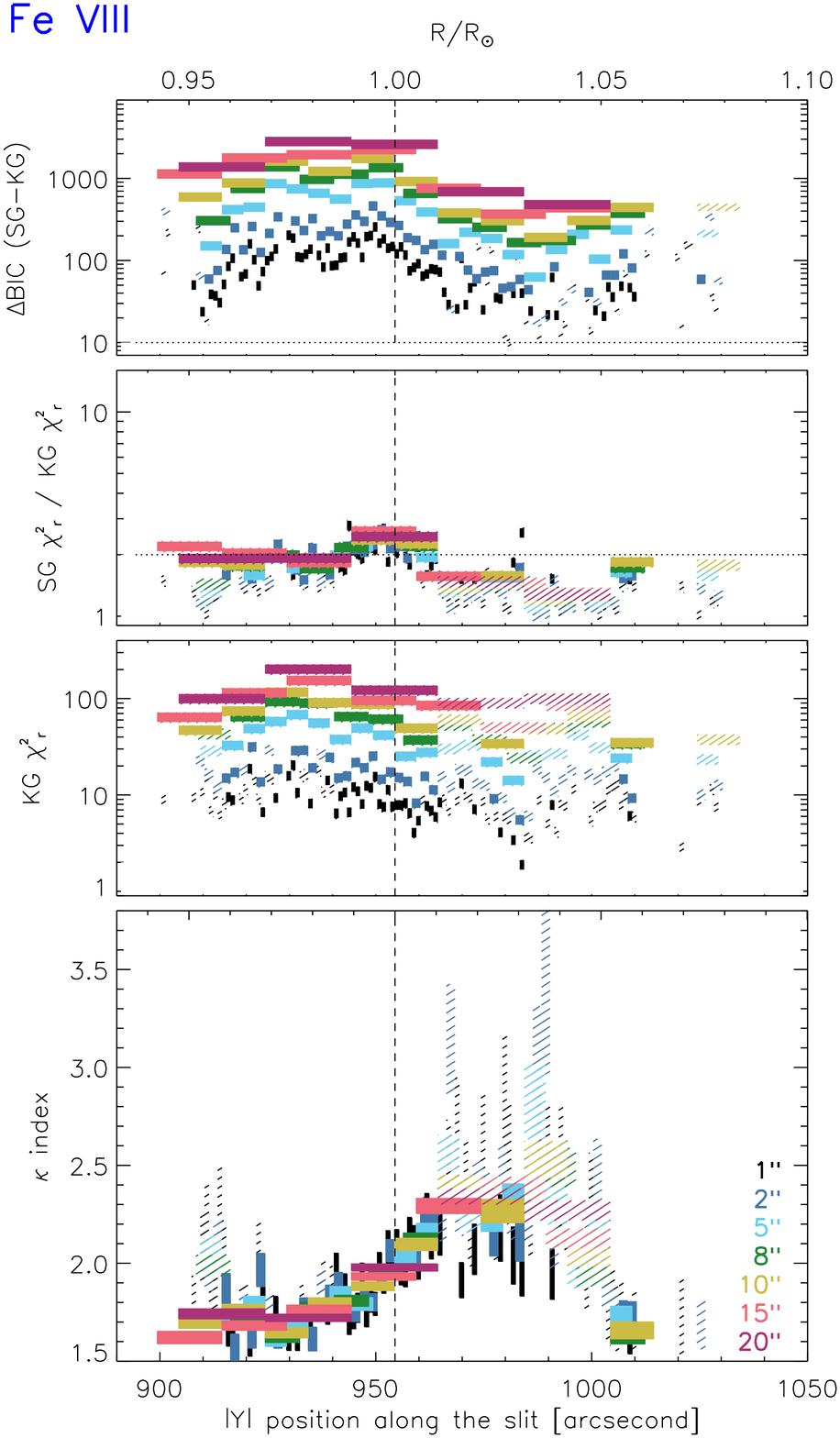}
\caption{A comparison of model goodness-of-fit parameters: $\Delta$BIC (SG-KG), ratio of $\chi_{\rm r}^{2}$ (SG/KG), and $\chi_{\rm r}^{2}$ (KG) using different spatial binning along $Y$ of $1\arcsec$, $2\arcsec$, $5\arcsec$, $8\arcsec$, $10\arcsec$, $15\arcsec$, and $20\arcsec$ for \ion{Si}{7} 275.361 \AA~{\it (left)} and \ion{Fe}{8} 186.599 \AA~{\it (right)}. Dashed boxes represent all lines fitted with a KG and solid boxes represent observations that satisfy the relaxed criterion. In the top four panels, the observations above all the dotted lines satisfy the full criteria. The vertical dashed line denotes the limb.}
\label{fig3}
\end{figure*}

\begin{figure*}[t]
\centering
\includegraphics[width=0.497\linewidth]{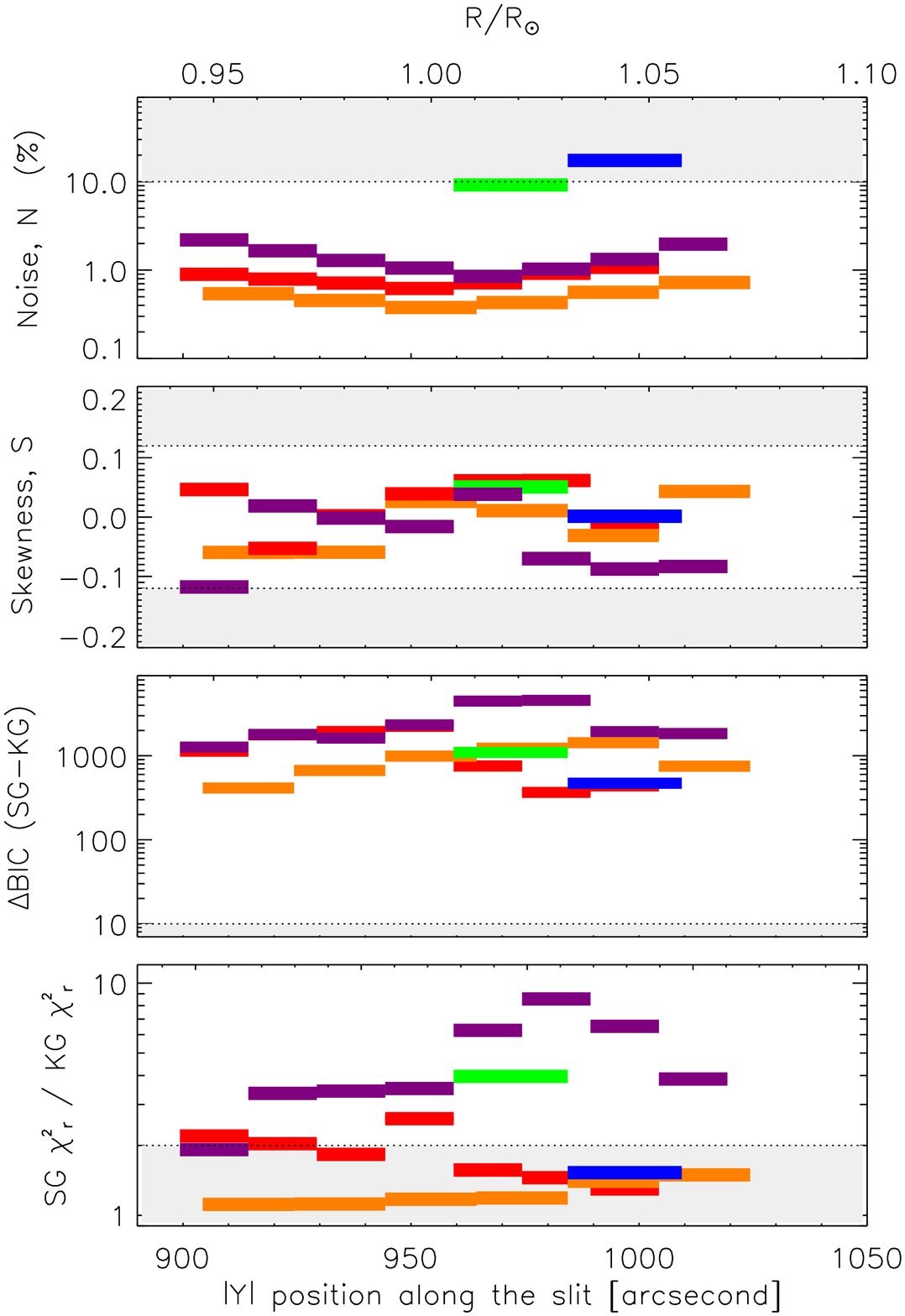}
\includegraphics[width=0.497\linewidth]{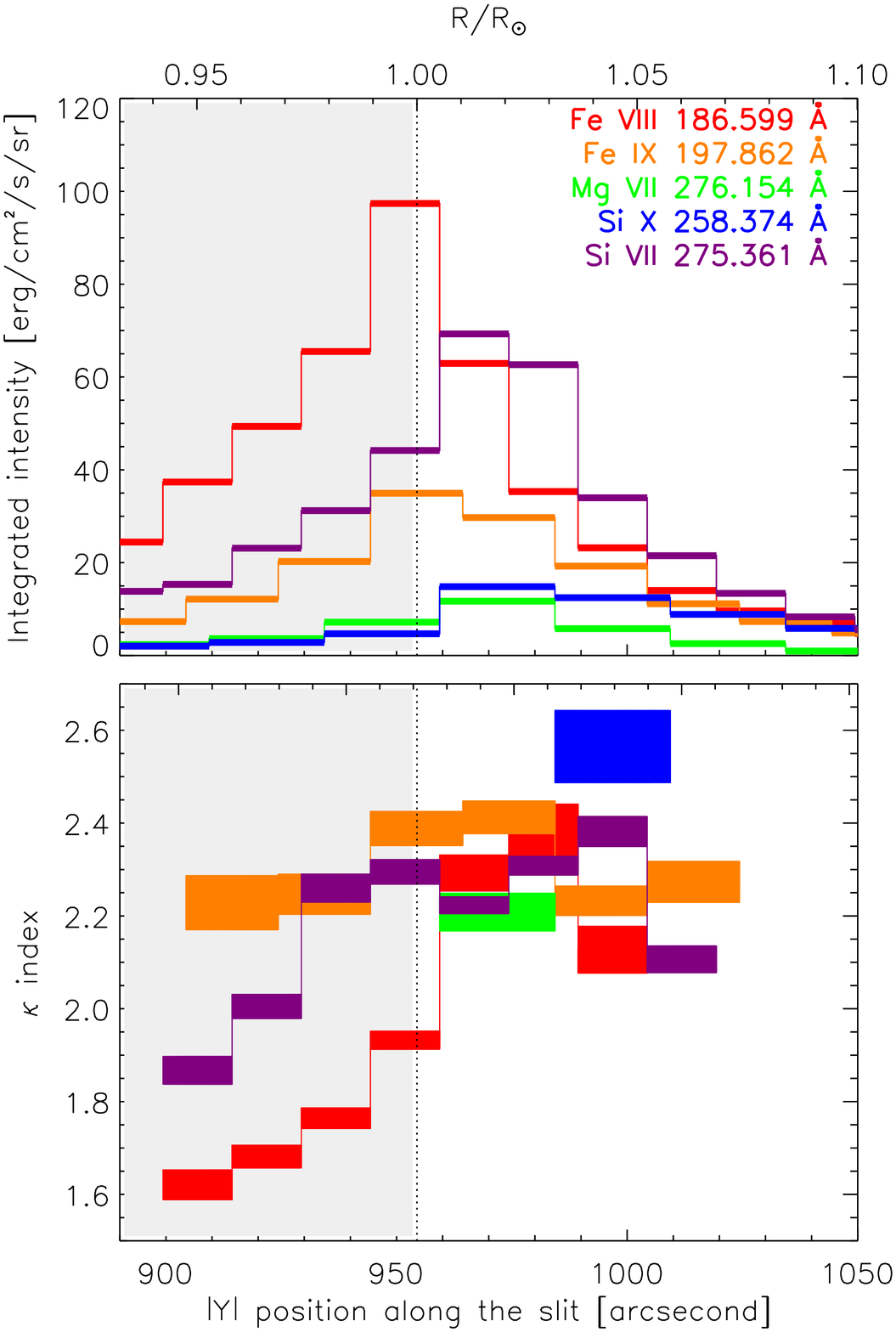}
\caption{{\it Left:} A comparison of the line and model goodness-of-fit parameters: noise $N$, skewness $S$, $\Delta$BIC (SG-KG), and ratio of $\chi_{\rm r}^{2}$ (SG/KG), for \ion{Fe}{8} 186.599 \AA~(red), \ion{Fe}{9} 197.862 \AA~(orange), \ion{Mg}{7} 276.154 \AA~(green), \ion{Si}{10} 258.374 \AA~(blue) and \ion{Si}{7} 275.361 \AA~(purple), for observations that satisfy the relaxed criterion. Lines that satisfy the full criterion lie within all white regions. {\it Right:} A comparison for each ion of the line integrated intensities {\it (top)} and $\kappa$ {\it (bottom)}. Solid rectangles show the $\kappa$ uncertainty and spatial binning. The grey region denotes altitudes below $1R_{\sun}$.}
\label{fig4}
\end{figure*}
Spectral lines suitable for analysis are free of blends and located far away from other lines. We require that the lines have an absolute value of skewness, $S$, (the third moment measuring asymmetry) lower than $~0.1$ and a ``noise'', $N$, less than $\approx10$\%, where $N$ is defined as the standard deviation of all indivdual line intensity errors divided by their intensities. This choice of criteria is discussed in \citet{2016A&A...590A..99J,2017ApJ...836...35J}. The lines suitable here for a profile analysis are: \ion{Si}{7}~275.361~\AA, \ion{Si}{10}~258.374~\AA, \ion{Mg}{7}~276.154~\AA, \ion{Fe}{8}~186.599~\AA~ and \ion{Fe}{9}~197.862~\AA. The main properties of each line are shown in Table \ref{tb1}. These lines have formation temperatures $T$ in units of K between $\log T={5.7-6.2}$ ($T\approx0.5-1.5$ MK) and comprise of three elements (Mg, Si, and Fe) with three different masses and charge-to-mass ratios in the range of $Q/M = 0.14-0.36$, where $Q$ is in units of $e$ and $M$ is the ion mass in atomic mass units. 

Each of our three model functions has a different number of free parameters: $\nu=4$ for SG, $5$ for KG, and $6$ for DG. In order to compare them to one another and determine the best model for each line profile, we use three different figures of merit. 
\begin{enumerate}
\item The fit residuals defined as $R_{i}=\frac{o_{i}-m_{i}}{\epsilon_{i}}$, where $o_{i}$ are the observed intensity values, $\epsilon_{i}$ are the observed intensity error values, and $m_{i}$ are the model values.
\item The reduced chi-squared $\chi_{r}^{2}=\frac{\chi^{2}}{\rm DOF}=\frac{1}{\rm DOF} \displaystyle \sum_{i}R_{i}^{2}$ from the weighted least-squares fit, where $\chi^{2}$ is the full chi-squared value. The degrees of freedom ${\rm DOF}=N - \nu$ where $N$ is the number of data points.
\item The Bayesian Information Criterion (or BIC test) \citep{schwarz1978} defined as ${\rm BIC}=-2\ln(L)+\nu \ln(N)$, where $L$ is the likelihood \citep[e.g., ][]{2017bmad.book.....H}. If the intensity errors are normally distributed then $-\ln(L)=\frac{1}{2}\sum_{i=1}^{N}\frac{(o_{i}-m_{i})^{2}}{\epsilon^{2}_{i}}+\frac{1}{2}\sum_{i=1}^{N}\ln(2\pi\epsilon_{i}^{2})$ \cite[e.g., ][]{1992nrfa.book.....P}. The $\Delta{\rm BIC}$, i.e. the difference between model $i$ and model $j$, can then be written as $\Delta {\rm BIC}={\rm BIC_{i}}-{\rm BIC_{j}}=\chi_{i}^{2}-\chi_{j}^{2}+\nu_{i} \ln(N)-\nu_{j} \ln(N)$.
 \end{enumerate}
The BIC test is useful as it explicitly accounts for the different number of free parameters in the model. Based on these tests, we define criteria that will allow us to state with confidence if a line profile is non-Gaussian, using the $\kappa$ index. The main criterion is that $\Delta {\rm BIC}={\rm SG\;BIC}-{\rm KG\;BIC}\ge 10$ \citep{WICS:WICS199}. In most cases $\Delta{\rm BIC}$ is very large ($\gtrsim100$). We also have confidence that the differences are significant when the SG $\chi_{\rm r}^{2}$ is at least twice as large as the KG $\chi_{\rm r}^{2}$. However, we will show in Section \ref{results} that $\chi_{\rm r}^{2}$ is often very large after binning the data, so that $\Delta$BIC is our prime method of model comparison. Also, there are cases where the line noise or skewness lies marginally outside the stated criteria, but such lines might still suitable for our analysis.

\section{Results and Uncertainties}\label{results}

Figure \ref{fig2} shows line fitting examples that met the criteria discussed above. The spectral line intensities are plotted using a log scale that show the wings clearly, since they have an intensity $1-2$ orders of magnitude smaller than the peak. All lines show broad wings to different degrees. For \ion{Si}{7} in particular, it is clear that a single Gaussian is unable to fit these broad wings. The smaller insert panels show the line peak using a linear scale, demonstrating that a single Gaussian is also unable to fit the peak in most cases. However, all models including KG, produce large values of $\chi_{\rm r}^{2} \approx 50-100$, since the intensity errors are small, in many cases only $0.1-1$\% of the intensity values. This produces residuals with values of $R=10$ (Figure \ref{fig2}). A likely reason for the relatively large $\chi_{\rm r}^{2}$ values, is that neither a KG, SG, nor DG distribution is an adequate model, particularly after we combine counts from many different plasma elements both along the line-of-sight in one pixel and over many pixels after binning. The poor $\chi_{\rm r}^{2}$ value then reflects the remaining discrepancy between the model and the physical distribution. We emphasize that our analysis detects non-Gaussian line profiles, and the kappa function is a convenient profile to detect such lines, but they are not necessarily kappa VDFs. 

Due to the large $\chi_{\rm r}^{2}$, $\Delta$BIC is primarily used to compare the different models, since it is a way of comparing different fits while controlling for the fact that a fit with more free parameters is bound to be better. However, the computation of $\Delta$BIC uses $\chi^{2}$, so it is important that the $\chi_{\rm r}^{2}$ are shown, and compared for different spatial binning. In Figure \ref{fig3}, we test different spatial binning of: $1\arcsec$, $2\arcsec$, $5\arcsec$, $8\arcsec$, $10\arcsec$, $15\arcsec$, and $20\arcsec$, using the most intense lines of \ion{Si}{7} and \ion{Fe}{8}. Figure \ref{fig3} shows that for smaller binning, the models do represent a more acceptable fit to the data. However for smaller binning, the KG $\chi_{\rm r}^{2}$ values are still $\gtrsim 3$ and often $\gtrsim 10$, even for $1\arcsec$ binning. More importantly, the spatial binning, and hence large $\chi_{\rm r}^{2}$, do not change the results of the $\Delta$BIC test,  nor does it significantly change the inferred values of $\kappa$. Thus, the spatial binning does not change the result that non-Gaussian spectral lines exist in this data or the results of our $\Delta$BIC test, even if kappa does not describe the underlying VDF correctly. For the comparison of different lines, we use the binning defined in Section \ref{data_analysis}, allowing: (1) an analysis over a slightly larger altitude range and (2) the study of weaker lines such as \ion{Mg}{7} and \ion{Si}{10}. 

A possible systematic effect is that the line broadening might be due to a flow component of the fast solar wind along the line of sight, but we can rule this out. At these heights the solar wind speed is $\approx 10$~$\mathrm{km\,s^{-1}}$ \citep{1999ApJ...511..481C}. At 200~\AA, a flow of 10~$\mathrm{km\,s^{-1}}$ corresponds to a Doppler shift of $\approx 0.007$~\AA. Assuming that the coronal hole is perpendicular to the center line of sight and reaches a maximum angle of roughly $30^{\circ}$ at the edge, the solar wind could, at most, shift the profiles by $0.002$~\AA. The observed broadening of the line wings relative to the SG profile is at least an order of magnitude larger. Thus, it is unlikely that solar wind flow alone is the cause of the observed non-Gaussian profiles. An earlier analysis of these coronal hole data by \citet{2012ApJ...753...36H} and \citet{2013ApJ...776...78H} show that instrument stray light is negligible at low altitudes.

We also fit the lines with another KG function (KG2), where the kappa part represents the instrumental profile and the Gaussian part represents the physical line profile \citep[as in ][]{2017ApJ...836...35J}. We use this function to test whether the non-Gaussian features of the line are an instrumental effect. As expected, the KG2 kappa part can account for the non-Gaussian features of the line, and we find that KG2 gives $\chi_{\rm r}^{2}$ and $\Delta$BIC values similar to KG. For the lines considered here, the instrumental width (FWHM) varies between $0.064-0.069$~\AA~over the range of $Y$ studied, and this varying width is accounted for during fitting. However, our use of the KG2 function shows that for a given $Y$ position, and hence for {\it fixed} instrumental values of width at that position, KG2 gives {\it different} instrumental values of $\kappa$ for each ion. This result implies that the non-Gaussian profiles are more likely physical, since otherwise the instrumental line shapes (and hence $\kappa$) would be the same at each $Y$ position, in contrast to our findings that they vary.

Figure \ref{fig4} compares the noise $N$ and skewness $S$ of all lines studied as well as the model goodness-of-fit indicators: $\Delta {\rm BIC}$ and the $\chi_{\rm r}^{2}$ ratio versus altitude, using the spatial binning given in Section \ref{data_analysis}. All of the lines suitable for study fall at low altitudes $>-1030 \arcsec$ ($\lesssim 1.1 R_{\sun}$), where the intensities are high. There are 23 altitudes where the lines are suitable for analysis (Table~\ref{tb1}). In every case for which the data were good enough to apply the analysis, we found non-Gaussian profiles. There were no cases in which SG provided a significantly better fit than KG. Figure~\ref{fig4} illustrates the derived $\kappa$ index versus altitude for each of the lines as we now discuss: 

\textit{\ion{Si}{7} 275.36 \AA~}is formed at $\log T\approx5.8$ and is the best line for this study. It has low noise with the background $2-3$ orders of magnitude below the peak intensity. Eight spatial bins were suitable for study between $0.98-1.1R_{\sun}$. The broad wings are clearly visible. In all cases, $\Delta {\rm BIC}\approx 1000$, indicates that the KG function better describes the line profile. We note that the DG function is also a better model than the SG. The $\kappa$ indices inferred from the KG fits are low, ranging from $\kappa\approx1.9-2.3$.

\textit{\ion{Fe}{8} 186.60 \AA~}is formed at $\log T \approx 5.7$ and it is more intense at low altitudes. However, the background is larger. Seven spatial bins were suitable for analysis. The fits were significantly non-Gaussian with $\kappa\approx1.6-2.3$. 

\textit{\ion{Fe}{9} 197.86 \AA~}is formed at $\log T \approx 6.0$. The spectral window containing \ion{Fe}{9} has a high background that could mask the low intensity wings, and \ion{Fe}{9} is less intense than \ion{Si}{7} or \ion{Fe}{8}. However, seven \ion{Fe}{9} spatial bins do satisfy the $\Delta {\rm BIC}$ difference test with values $>20$. Considering the positions satisfying the relaxed criterion we find $\kappa\approx2.0-2.3$. 

\textit{\ion{Mg}{7} 276.15 \AA~}is formed at $\log T \approx 5.8$ and has a low intensity. The \ion{Mg}{7} line was suitable for analysis one spatial bin. We found $\kappa\approx2.1$. 
	
\textit{\ion{Si}{10} 258.37 \AA~}is formed at $\log T \approx 6.2$. The lines from this ion were not very intense and \ion{Si}{10} was only suitable for analysis at one spatial bin. We found a non-Gaussian profile with $\kappa\approx2.5$. 

\section{Discussion}
We find non-Gaussian line profiles low in a coronal hole, which are better represented by kappa-distributions with $\kappa < 2.6$. The inferred value of $\kappa$ increases moving radially outward until the observations reach the solar limb, as can be seen for both \ion{Si}{7} and \ion{Fe}{8} before $\approx 1 R_{\odot}$. Above this altitude, \ion{Si}{7} $\kappa$ remains approximately constant at values of $\kappa\approx2.1-2.3$. The other lines have similar $\kappa$ values. The initial increase of $\kappa$ in the on-disk data may be due to flows along the line of sight. Above the limb, the line of sight is nearly perpendicular to the magnetic field so that flows are greatly reduced. In principle, flows from spicules and low-lying loops could affect the profiles in the off-limb data, but $\kappa$ becomes approximately constant above the limb, suggesting that this is not an important effect. Nevertheless, this should be confirmed in future work by studying lines at larger altitudes. 

There appears to be a weak $\kappa$ dependence with temperature, with $\kappa$ increasing slightly with increasing temperature. However, the changes are small since the entire off-limb $\kappa$ range is only $\approx 1.9-2.5$. If this trend is real, one possible explanation is that it is caused by different structures along the line-of-sight. Cooler lines are likely to be emitted by the coronal hole plasma, whereas warmer lines could be contaminated by quiet Sun emission. As the quiet Sun is denser than the coronal hole, there may be more collisions driving those lines toward a Maxwellian.

The data were not sufficient to identify the physical mechanism generating the profiles, although our analysis clearly shows non-Gaussian line profiles at the base of the fast solar wind. One explanation is fluid motions such as turbulence or waves. Alternatively, we might be observing non-thermal ion VDFs, possibly from ion heating. Hence, the profiles may be due to macroscopic fluid motions, microscopic ion motions, or both. 

One way to distinguish fluid motions from ion motions is to study ions of different masses $M$ or charge-to-mass ratios, $Q/M$. Fluid motions might be expected to affect all ions in the same way, but the ion VDFs may differ due to heating that depends on $M$ or $Q/M$. Similar techniques have been used to separate thermal from non-thermal broadening in spectral lines \citep{Tu:ApJ:1998, Landi:ApJ:2009, 2013ApJ...776...78H,2013ApJ...763..106H}. Alternatively, processes such as the cascade of MHD waves towards smaller scales might be expected to produce more efficient ion acceleration, and hence lower $\kappa$ indices, for ions with lower cyclotron frequencies $\Omega_{c,i}$, since $\Omega_{c,i}=(Q/M) \Omega_{c,p}$ \citep[e.g., see the recent review by][]{2017SSRv..212.1107K}, where $\Omega_{c,p}$ is the proton cyclotron frequency.

Here we find no significant trend in $\kappa$ with $M$ or $Q/M$, but it is also likely that any trend in $\kappa$ with $M$ or $Q/M$ is hard to identify because of the measurement uncertainties, the small number of lines, and the limited range of altitudes. Moreover, we note that Si and Mg are lighter than Fe, so any real effects from their VDF due to ion acceleration might be more evident in their line profile because they should have a larger velocity.

Finally, at low altitudes, the Doppler motions are observed perpendicular to the magnetic field, which could be suggestive of, e.g., Alfv\'enic fluctuations. Evidence of Alfv\'en wave dissipation via non-thermal broadening was presented in \citet{2013ApJ...776...78H} for these data. While wave dissipation and turbulence are normally diagnosed from line broadening \citep[e.g., ][]{2017PhRvL.118o5101K}, non-Gaussian VDFs might provide evidence of turbulent intermittency in the corona, a property of solar wind turbulence \citep[cf., ][]{2013SSRv..178..101A}. Clearly additional observational work is needed to better constrain the physical mechanism(s) generating the observed non-Gaussian line profiles.

\section{Summary}
We have reported the detection of non-Gaussian EUV spectral line shapes at the base of the fast solar wind ($< 1.1$~$R_{\sun}$). We quantified the non-Gaussian properties of these lines in terms of $\kappa$, which were found to range off-limb between $\approx 1.9-2.5$. The cause of these non-Gaussian line profiles may be (a) non-Maxwellian ion VDFs at the base of fast solar wind, (b) fluid motions such as non-Gaussian turbulent fluctuations or non-uniform wave motions, or (c) some combination of both. The findings are a timely precursor to future observations with the {\it Parker Solar Probe} \citep{2016SSRv..204....7F}, that will study ion VDFs \textit{in situ} as close as $\approx8.5 R_{\sun}$, and hence, test whether non-Gaussian VDFs are indeed formed very close to the Sun.

\acknowledgments
NLSJ \& LF gratefully acknowledge the financial support provided by the STFC Consolidated Grant ST/L000741/1, and the grant awarded by the Principal's Early Career Mobility Scheme (University of Glasgow). MH \& DWS were supported by the NASA Living with a Star Program grant NNX15AB71G and by the NSF Division of Atmospheric and Geospace Sciences SHINE program grant AGS-1459247. Hinode is a Japanese mission developed and launched by ISAS/JAXA, collaborating with NAOJ as a domestic partner, NASA and UKSA as international partners. Scientific operation of the Hinode mission is conducted by the Hinode science team organized at ISAS/JAXA. This team mainly consists of scientists from institutes in the partner countries. Support for the post-launch operation is provided by JAXA and NAOJ (Japan), UKSA (U.K.), NASA, ESA, and NSC (Norway).

%\bibliographystyle{aasjournal}
%\bibliography{ms1}

\begin{thebibliography}{}
\expandafter\ifx\csname natexlab\endcsname\relax\def\natexlab#1{#1}\fi
\providecommand{\url}[1]{\href{#1}{#1}}

\bibitem[{{Alexandrova} {et~al.}(2013){Alexandrova}, {Chen}, {Sorriso-Valvo},
  {Horbury}, \& {Bale}}]{2013SSRv..178..101A}
{Alexandrova}, O., {Chen}, C.~H.~K., {Sorriso-Valvo}, L., {Horbury}, T.~S., \&
  {Bale}, S.~D. 2013, \ssr, 178, 101

\bibitem[{{Bruno} \& {Carbone}(2005)}]{2005LRSP....2....4B}
{Bruno}, R., \& {Carbone}, V. 2005, Living Reviews in Solar Physics, 2, 4

\bibitem[{Chae {et~al.}(1998)Chae, Sch\"uhle, \& Lemaire}]{Chae:ApJ:1998}
Chae, J., Sch\"uhle, U., \& Lemaire, P. 1998, ApJ, 505, 957

\bibitem[{Collier {et~al.}(1996)Collier, Hamilton, Gloeckler, Bochsler, \&
  Sheldon}]{GRL:GRL9092}
Collier, M.~R., Hamilton, D.~C., Gloeckler, G., Bochsler, P., \& Sheldon, R.~B.
  1996, Geophysical Research Letters, 23, 1191.
\newblock \url{http://dx.doi.org/10.1029/96GL00621}

\bibitem[{{Cranmer}(2002)}]{2002ESASP.508..361C}
{Cranmer}, S.~R. 2002, in ESA Special Publication, Vol. 508, From Solar Min to
  Max: Half a Solar Cycle with SOHO, ed. A.~{Wilson}, 361--366

\bibitem[{{Cranmer}(2009)}]{2009LRSP....6....3C}
{Cranmer}, S.~R. 2009, Living Reviews in Solar Physics, 6, 3

\bibitem[{{Cranmer} {et~al.}(1999){Cranmer}, {Kohl}, {Noci}, {Antonucci},
  {Tondello}, {Huber}, {Strachan}, {Panasyuk}, {Gardner}, {Romoli}, {Fineschi},
  {Dobrzycka}, {Raymond}, {Nicolosi}, {Siegmund}, {Spadaro}, {Benna},
  {Ciaravella}, {Giordano}, {Habbal}, {Karovska}, {Li}, {Martin}, {Michels},
  {Modigliani}, {Naletto}, {O'Neal}, {Pernechele}, {Poletto}, {Smith}, \&
  {Suleiman}}]{1999ApJ...511..481C}
{Cranmer}, S.~R., {Kohl}, J.~L., {Noci}, G., {et~al.} 1999, \apj, 511, 481

\bibitem[{{Culhane} {et~al.}(2007){Culhane}, {Harra}, {James}, {Al-Janabi},
  {Bradley}, {Chaudry}, {Rees}, {Tandy}, {Thomas}, {Whillock}, {Winter},
  {Doschek}, {Korendyke}, {Brown}, {Myers}, {Mariska}, {Seely}, {Lang}, {Kent},
  {Shaughnessy}, {Young}, {Simnett}, {Castelli}, {Mahmoud}, {Mapson-Menard},
  {Probyn}, {Thomas}, {Davila}, {Dere}, {Windt}, {Shea}, {Hagood}, {Moye},
  {Hara}, {Watanabe}, {Matsuzaki}, {Kosugi}, {Hansteen}, \&
  {Wikstol}}]{2007SoPh..243...19C}
{Culhane}, J.~L., {Harra}, L.~K., {James}, A.~M., {et~al.} 2007, \solphys, 243,
  19

\bibitem[{{De Pontieu} {et~al.}(2014){De Pontieu}, {Title}, {Lemen}, {Kushner},
  {Akin}, {Allard}, {Berger}, {Boerner}, {Cheung}, {Chou}, {Drake}, {Duncan},
  {Freeland}, {Heyman}, {Hoffman}, {Hurlburt}, {Lindgren}, {Mathur}, {Rehse},
  {Sabolish}, {Seguin}, {Schrijver}, {Tarbell}, {W{\"u}lser}, {Wolfson},
  {Yanari}, {Mudge}, {Nguyen-Phuc}, {Timmons}, {van Bezooijen}, {Weingrod},
  {Brookner}, {Butcher}, {Dougherty}, {Eder}, {Knagenhjelm}, {Larsen},
  {Mansir}, {Phan}, {Boyle}, {Cheimets}, {DeLuca}, {Golub}, {Gates}, {Hertz},
  {McKillop}, {Park}, {Perry}, {Podgorski}, {Reeves}, {Saar}, {Testa}, {Tian},
  {Weber}, {Dunn}, {Eccles}, {Jaeggli}, {Kankelborg}, {Mashburn}, {Pust},
  {Springer}, {Carvalho}, {Kleint}, {Marmie}, {Mazmanian}, {Pereira}, {Sawyer},
  {Strong}, {Worden}, {Carlsson}, {Hansteen}, {Leenaarts}, {Wiesmann},
  {Aloise}, {Chu}, {Bush}, {Scherrer}, {Brekke}, {Martinez-Sykora}, {Lites},
  {McIntosh}, {Uitenbroek}, {Okamoto}, {Gummin}, {Auker}, {Jerram}, {Pool}, \&
  {Waltham}}]{2014SoPh..289.2733D}
{De Pontieu}, B., {Title}, A.~M., {Lemen}, J.~R., {et~al.} 2014, \solphys, 289,
  2733

\bibitem[{{Delaboudini{\`e}re} {et~al.}(1995){Delaboudini{\`e}re}, {Artzner},
  {Brunaud}, {Gabriel}, {Hochedez}, {Millier}, {Song}, {Au}, {Dere}, {Howard},
  {Kreplin}, {Michels}, {Moses}, {Defise}, {Jamar}, {Rochus}, {Chauvineau},
  {Marioge}, {Catura}, {Lemen}, {Shing}, {Stern}, {Gurman}, {Neupert},
  {Maucherat}, {Clette}, {Cugnon}, \& {van Dessel}}]{1995SoPh..162..291D}
{Delaboudini{\`e}re}, J.-P., {Artzner}, G.~E., {Brunaud}, J., {et~al.} 1995,
  \solphys, 162, 291

\bibitem[{{Dere} {et~al.}(1997){Dere}, {Landi}, {Mason}, {Monsignori Fossi}, \&
  {Young}}]{1997A&AS..125..149D}
{Dere}, K.~P., {Landi}, E., {Mason}, H.~E., {Monsignori Fossi}, B.~C., \&
  {Young}, P.~R. 1997, \aaps, 125, 149

\bibitem[{{Domingo} {et~al.}(1995){Domingo}, {Fleck}, \&
  {Poland}}]{1995SSRv...72...81D}
{Domingo}, V., {Fleck}, B., \& {Poland}, A.~I. 1995, \ssr, 72, 81

\bibitem[{{Dud{\'{\i}}k} {et~al.}(2017){Dud{\'{\i}}k}, {Polito}, {Dzif{\v
  c}{\'a}kov{\'a}}, {Del Zanna}, \& {Testa}}]{2017ApJ...842...19D}
{Dud{\'{\i}}k}, J., {Polito}, V., {Dzif{\v c}{\'a}kov{\'a}}, E., {Del Zanna},
  G., \& {Testa}, P. 2017, \apj, 842, 19

\bibitem[{{Fox} {et~al.}(2016){Fox}, {Velli}, {Bale}, {Decker}, {Driesman},
  {Howard}, {Kasper}, {Kinnison}, {Kusterer}, {Lario}, {Lockwood}, {McComas},
  {Raouafi}, \& {Szabo}}]{2016SSRv..204....7F}
{Fox}, N.~J., {Velli}, M.~C., {Bale}, S.~D., {et~al.} 2016, \ssr, 204, 7

\bibitem[{{Gloeckler} {et~al.}(1995){Gloeckler}, {Balsiger}, {B{\"u}rgi},
  {Bochsler}, {Fisk}, {Galvin}, {Geiss}, {Gliem}, {Hamilton}, {Holzer},
  {Hovestadt}, {Ipavich}, {Kirsch}, {Lundgren}, {Ogilvie}, {Sheldon}, \&
  {Wilken}}]{1995SSRv...71...79G}
{Gloeckler}, G., {Balsiger}, H., {B{\"u}rgi}, A., {et~al.} 1995, \ssr, 71, 79

\bibitem[{{Hahn} {et~al.}(2012){Hahn}, {Landi}, \&
  {Savin}}]{2012ApJ...753...36H}
{Hahn}, M., {Landi}, E., \& {Savin}, D.~W. 2012, \apj, 753, 36

\bibitem[{{Hahn} \& {Savin}(2013{\natexlab{a}})}]{2013ApJ...776...78H}
{Hahn}, M., \& {Savin}, D.~W. 2013{\natexlab{a}}, \apj, 776, 78

\bibitem[{{Hahn} \& {Savin}(2013{\natexlab{b}})}]{2013ApJ...763..106H}
---. 2013{\natexlab{b}}, \apj, 763, 106

\bibitem[{{Hilbe} {et~al.}(2017){Hilbe}, {de Souza}, \&
  {Ishida}}]{2017bmad.book.....H}
{Hilbe}, J.~M., {de Souza}, R.~S., \& {Ishida}, E.~E.~O. 2017, {Bayesian Models
  for Astrophysical Data Using R, JAGS, Python, and Stan},
  doi:10.1017/CBO9781316459515

\bibitem[{{Jeffrey} {et~al.}(2016){Jeffrey}, {Fletcher}, \&
  {Labrosse}}]{2016A&A...590A..99J}
{Jeffrey}, N.~L.~S., {Fletcher}, L., \& {Labrosse}, N. 2016, \aap, 590, A99

\bibitem[{{Jeffrey} {et~al.}(2017){Jeffrey}, {Fletcher}, \&
  {Labrosse}}]{2017ApJ...836...35J}
---. 2017, \apj, 836, 35

\bibitem[{Kasper {et~al.}(2008)Kasper, Lazarus, \& Gary}]{Kasper:PRL:2008}
Kasper, J.~C., Lazarus, A.~J., \& Gary, S.~P. 2008, Phys.\ Rev.\ Lett., 101,
  261103

\bibitem[{{Klein} \& {Dalla}(2017)}]{2017SSRv..212.1107K}
{Klein}, K.-L., \& {Dalla}, S. 2017, \ssr, 212, 1107

\bibitem[{{Klimchuk} {et~al.}(2016){Klimchuk}, {Patsourakos}, \&
  {Tripathi}}]{2016SoPh..291...55K}
{Klimchuk}, J.~A., {Patsourakos}, S., \& {Tripathi}, D. 2016, \solphys, 291, 55

\bibitem[{{Kohl} {et~al.}(1995){Kohl}, {Esser}, {Gardner}, {Habbal},
  {Daigneau}, {Dennis}, {Nystrom}, {Panasyuk}, {Raymond}, {Smith}, {Strachan},
  {van Ballegooijen}, {Noci}, {Fineschi}, {Romoli}, {Ciaravella}, {Modigliani},
  {Huber}, {Antonucci}, {Benna}, {Giordano}, {Tondello}, {Nicolosi}, {Naletto},
  {Pernechele}, {Spadaro}, {Poletto}, {Livi}, {von der L{\"u}he}, {Geiss},
  {Timothy}, {Gloeckler}, {Allegra}, {Basile}, {Brusa}, {Wood}, {Siegmund},
  {Fowler}, {Fisher}, \& {Jhabvala}}]{1995SoPh..162..313K}
{Kohl}, J.~L., {Esser}, R., {Gardner}, L.~D., {et~al.} 1995, \solphys, 162, 313

\bibitem[{Kohl {et~al.}(1998)Kohl, Noci, Antonucci, Tondello, Huber, Cranmer,
  Strachan, Panasyuk, Gardner, Romoli, Fineschi, Dobrzycka, Raymond, Nicolosi,
  Siegmund, Spadaro, Benna, Ciaravella, Giordano, Habbal, Karovska, Li, Martin,
  Michels, Modigliani, Naletto, {O'Neal}, Pernechele, Poletto, Smith, \&
  Suleiman}]{Kohl:ApJ:1998}
Kohl, J.~L., Noci, G., Antonucci, E., {et~al.} 1998, ApJ, 501, L127

\bibitem[{{Kontar} {et~al.}(2017){Kontar}, {Perez}, {Harra}, {Kuznetsov},
  {Emslie}, {Jeffrey}, {Bian}, \& {Dennis}}]{2017PhRvL.118o5101K}
{Kontar}, E.~P., {Perez}, J.~E., {Harra}, L.~K., {et~al.} 2017, Physical Review
  Letters, 118, 155101

\bibitem[{{Kosugi} {et~al.}(2007){Kosugi}, {Matsuzaki}, {Sakao}, {Shimizu},
  {Sone}, {Tachikawa}, {Hashimoto}, {Minesugi}, {Ohnishi}, {Yamada}, {Tsuneta},
  {Hara}, {Ichimoto}, {Suematsu}, {Shimojo}, {Watanabe}, {Shimada}, {Davis},
  {Hill}, {Owens}, {Title}, {Culhane}, {Harra}, {Doschek}, \&
  {Golub}}]{2007SoPh..243....3K}
{Kosugi}, T., {Matsuzaki}, K., {Sakao}, T., {et~al.} 2007, \solphys, 243, 3

\bibitem[{Landi \& Cranmer(2009)}]{Landi:ApJ:2009}
Landi, E., \& Cranmer, S.~R. 2009, ApJ, 691, 794

\bibitem[{{Landi} {et~al.}(2013){Landi}, {Young}, {Dere}, {Del Zanna}, \&
  {Mason}}]{2013ApJ...763...86L}
{Landi}, E., {Young}, P.~R., {Dere}, K.~P., {Del Zanna}, G., \& {Mason}, H.~E.
  2013, \apj, 763, 86

\bibitem[{{Lee} {et~al.}(2013){Lee}, {Williams}, \&
  {Lapenta}}]{2013arXiv1305.2939L}
{Lee}, E., {Williams}, D.~R., \& {Lapenta}, G. 2013, ArXiv e-prints,
  arXiv:1305.2939

\bibitem[{{Livadiotis} \& {McComas}(2009)}]{2009JGRA..11411105L}
{Livadiotis}, G., \& {McComas}, D.~J. 2009, Journal of Geophysical Research
  (Space Physics), 114, 11105

\bibitem[{{Marandet} {et~al.}(2004){Marandet}, {Capes}, {Godbert-Mouret},
  {Koubiti}, \& {Stamm}}]{2004physics..12091M}
{Marandet}, Y., {Capes}, H., {Godbert-Mouret}, L., {Koubiti}, M., \& {Stamm},
  R. 2004, ArXiv Physics e-prints, physics/0412091

\bibitem[{{Marandet} \& {Dufty}(2006)}]{2006CoPP...46..672M}
{Marandet}, Y., \& {Dufty}, J.~W. 2006, Contributions to Plasma Physics, 46,
  672

\bibitem[{{Marsch}(2006)}]{2006LRSP....3....1M}
{Marsch}, E. 2006, Living Reviews in Solar Physics, 3, 1

\bibitem[{Neath \& Cavanaugh(2012)}]{WICS:WICS199}
Neath, A.~A., \& Cavanaugh, J.~E. 2012, Wiley Interdisciplinary Reviews:
  Computational Statistics, 4, 199.
\newblock \url{http://dx.doi.org/10.1002/wics.199}

\bibitem[{{Olbert}(1968)}]{1968ASSL...10..641O}
{Olbert}, S. 1968, in Astrophysics and Space Science Library, Vol.~10, Physics
  of the Magnetosphere, ed. R.~D.~L. {Carovillano} \& J.~F. {McClay}, 641

\bibitem[{{Press} {et~al.}(1992){Press}, {Teukolsky}, {Vetterling}, \&
  {Flannery}}]{1992nrfa.book.....P}
{Press}, W.~H., {Teukolsky}, S.~A., {Vetterling}, W.~T., \& {Flannery}, B.~P.
  1992, {Numerical recipes in FORTRAN. The art of scientific computing}

\bibitem[{Schwarz(1978)}]{schwarz1978}
Schwarz, G. 1978, Ann. Statist., 6, 461.
\newblock \url{http://dx.doi.org/10.1214/aos/1176344136}

\bibitem[{{Tu} \& {Marsch}(1995)}]{1995SSRv...73....1T}
{Tu}, C.-Y., \& {Marsch}, E. 1995, \ssr, 73, 1

\bibitem[{Tu {et~al.}(1998)Tu, Marsch, Wilhelm, \& Curdt}]{Tu:ApJ:1998}
Tu, C.-Y., Marsch, E., Wilhelm, K., \& Curdt, W. 1998, ApJ, 503

\bibitem[{{Vasyliunas}(1968)}]{1968JGR....73.2839V}
{Vasyliunas}, V.~M. 1968, \jgr, 73, 2839

\bibitem[{{Wilhelm} {et~al.}(1995){Wilhelm}, {Curdt}, {Marsch}, {Sch{\"u}hle},
  {Lemaire}, {Gabriel}, {Vial}, {Grewing}, {Huber}, {Jordan}, {Poland},
  {Thomas}, {K{\"u}hne}, {Timothy}, {Hassler}, \&
  {Siegmund}}]{1995SoPh..162..189W}
{Wilhelm}, K., {Curdt}, W., {Marsch}, E., {et~al.} 1995, \solphys, 162, 189

\end{thebibliography}

\end{document}